\documentclass[aps,a4paper,12pt]{revtex4-1}
\def\e{{\rm e}}
\def\a{\alpha}
\def\Erf{{\rm Erf}}
\usepackage[english]{babel}
\usepackage{amssymb,ulem}
\usepackage{amsmath}
\usepackage{indentfirst}
\usepackage{graphicx}
\usepackage{graphics}
\usepackage{epsfig}
\usepackage{subfigure}
\usepackage{multirow}
\usepackage{color}

\newcommand{\mathsym}[1]{{}}

\def\e{{\mathrm{e}}}

\def\betab{b}

\def\d{{\rm d}}
\def\e{{\rm e}}
\def\a{\alpha}
\def\Erf{{\rm Erf}}
\def\bfr{{\bf r}}
\def\bfp{{\bf p}}
\def\bfalpha{\boldsymbol{\alpha}}

\def\la{\lambda}
\def\b{\beta}
\def\aa{a}
\def\bb{b}

\begin{document}
\title{Quark meson coupling model with $su(3)$ symmetry  within  the Bogoliubov independent quark model of the nucleon}
\author{
Prafulla K. Panda\\
{\it\small Department of Physics, Utkal University, Bhubaneswar, 751004, India }\\
Constan\c{c}a Provid\^encia\\
{\it \small CFisUC, Departamento de F\'\i sica, Universidade de Coimbra,}\\
{\it \small  P-3004-516 Coimbra, Portugal}\\
Steven A.\ Moszkowski\\
{\sl \small UCLA, Los Angeles, CA 90095, USA} \\
Henrik Bohr\\
{\it \small Department of Physics, B.307, Danish Technical
University,}\\{ \it \small DK-2800 Lyngby, Denmark}\\
Jo\~ao da Provid\^encia\\
{\it \small CFisUC, Departamento de F\'\i sica, Universidade de Coimbra,}\\
{\it \small  P-3004-516 Coimbra, Portugal}
}
%\date{}

\begin{abstract}
We generalize the Bogoliubov quark-meson coupling model to also
include hyperons. The hyperon-$\sigma$-meson couplings are fixed by
the model and the  hyperon-$\omega$-meson couplings are fitted to the
hypreon potentials in symmetric nuclear matter. The present
model  predicts neutron stars with masses above 2$M_\odot$ and the
radius of a 1.4$M_\odot$ star equal to 13.83 km.
\end{abstract}
\maketitle
\section{Introduction}
The properties of nuclear matter has been an area of interest for the past few decades. Such studies
are quite important in nuclear physics in the context of nucleon-nucleon $(NN)$ interaction, structure and 
properties of finite nuclei, dynamics of heavy ion collisions, nuclear-astrophysics and also particle physics. 
The relativistic mean field (RMF) models \cite{walecka,boguta,muller,typel} represent
the $NN$ interactions through the coupling of nucleons with isoscalar
scalar mesons, isoscalar vector mesons, isovector vector mesons, and the photon
quanta besides the self- and cross-interactions  \cite{boguta,muller} among these
mesons, or density dependent couplings \cite{typel}.

In the above RMF models,  the nucleons are treated as structureless point objects.
However, incorporation of structure of nucleon with meson couplings at the
basic quark level in the study of saturation properties of nuclear
matter can provide new insight. Having this in mind, there have been several
attempts based on the MIT bag model  \cite{guichon} and on the Nambu-Jona-Lasinio
(NJL) model \cite{bentz} to address the nucleon structure.
Using such quark-meson coupling (QMC) models, the
nuclear equation of state (EOS) has also been constructed
and properties of nuclear matter have been studied
in great detail in a series of works by Guichon, Saito and Thomas \cite{guichon, guichon1,
guichon2} and by others \cite{qmcPanda, qmcPanda1, qmcPanda2,qmcPanda3}.
Recently the modified quark-meson coupling model which is based on
confining relativistic independent quark potential model rather
than a bag to describe the baryon structure in vacuum, has been extensively applied for the
study of the bulk properties of both symmetric as well as
asymmetric nuclear matter \cite{batista,barik1,barik2}.

The aim of the present work is to obtain a phenomenological description of hadronic matter 
including hyperonic degrees of freedom, in the spirit of the QMC approach, combined with the Bogoliubov
model \cite{bogolubov} for the description of the quark dynamics in
the nucleon \cite{bohr}. We will refer to the present model the
Bogoliubov-QMC model.  In \cite{bohr},  symmetric and asymmetric nuclear matter at saturation density
have been successfully described. In the present study we will
generalize the model in order to include hyperons and will study the structure of neutron stars within this model.

\section{Equation of state}
The independent quark model  of the nucleon proposed by Bogoliubov \cite{bogolubov}
is described by the Hamiltonian
\begin{equation}
\label{h_D}
h_D=-i\bfalpha\cdot\nabla+\beta\left(\kappa|\bfr|+m-g^q_\sigma
  \sigma\right),
\end{equation}
where $m$ is the current quark mass, $\beta$ and the components $\alpha_x,\alpha_y,\alpha_z$ of
$\bfalpha$ are Dirac matrices, $\sigma$ denotes the external scalar
field, $g^q_\sigma$ denotes the coupling of the quark to the $\sigma$ field and $\kappa$
denotes the string tension. The constituent quark mass is obtained by solving the Dirac equation,
\begin{equation}
\left[\bfalpha\cdot\bfp+\beta(\kappa|\bfr|+m-g^q_\sigma\sigma)\right]\psi_q=\varepsilon_q\psi_q.
\end{equation}
The current quark $m$ is taken to be
$m=0$ for $u,d$ quarks because their constituent mass
is assumed to be determined exclusively by the value of $\kappa$.
For the $s$-quark the value $m=232.633$~MeV has
been chosen in order to reproduce the $\Lambda$-hyperon mass 1130 MeV
in the vacuum, with the same
value of $\kappa$ which has been considered for quarks $u,~d$.

The eigenvalues of $h_D$ are obtained by a scale transformation from the eigenvalues of
\begin{equation*}h_{D0}=-i\bfalpha\cdot\nabla+\beta\left(|\bfr|-\aa\right).
\end{equation*}
which are determined by considering its square,
\begin{equation}\label{2D0} h_{D0}^2=-\nabla^2+(|\bfr|-
a)^2+i\beta\bfalpha\cdot{\bfr\over|\bfr|},\end{equation}
and looking for
the respective eigenvalues versus $a$. For $a=0$ the groundstate
eigenvalue of $h^2_{D0}$ reads $2.6402$ in units of $\kappa$ \cite{bohr}.
%Notice that $h_{D0}$, $a$, and also $r$ in (\ref{2D0}) are
%dimensionless.
$\kappa$ takes the value
$$\kappa={313^2\over2.6402}~{\rm MeV}^2=37106.9317~{\rm MeV}^2,$$
so as to reproduce the constituent mass 313 MeV of quarks $u,~d$ in vacuum.
For completeness sake, we describe in the Appendix the procedure followed in \cite{bohr} to determine
variationally the groundstate wave function of  $h^2_{D0}$.

We have found that in the interval $-1.25<a<2.4$, that covers the
range of densities we will consider, we may express the ground state energy, $m(\kappa,\aa),$  of $h_{D0}$,
with sufficient accuracy, as
\begin{eqnarray}&&\nonumber {(m(\kappa,\aa))^2\over\kappa}= 2.64022 - 2.3644 a + 0.76534 a^2 - 0.0468815  a^3 - 0.0131333 a^4 - 0.00323908 a^5\\&&  + 0.00117542 a^6 .\label{quarkmass}
\end{eqnarray}
%We have ensured that the interval  $-1.25<a<2.4$ is adequate for the range of densities we have considered.
We take $a=g^q_\sigma\sigma/\sqrt{\kappa}$ for quarks $u,d$ because, in vacuum, the constituent mass 313 MeV of these
quarks is described with $a=0$ and $\kappa=37106.9317$ MeV$^2$. For the quark $s$,
$a=a_s=%-1.2
-1.2455+g^q_\sigma\sigma/\sqrt{\kappa}$ reproduces the  vacuum  constituent mass
504 MeV of this quark.

Thus, the constituent mass of quarks $u,d$ is
\begin{eqnarray*}&&m_u=m_d=\sqrt\kappa\left( 2.64022 - 2.3644 a + 0.76534 a^2 - 0.0468815  a^3 - 0.0131333 a^4 -
0.00323908 a^5\right.\\&&\left.  + 0.00117542 a^6 \right)^{1/2},\end{eqnarray*}
where   $a=a_u=a_d=g^q_\sigma\sigma/\sqrt{\kappa}$, and the constituent mass of quarks $s$ is
\begin{eqnarray*}
&&m_s=\sqrt\kappa\left( 2.64022 - 2.3644 a + 0.76534 a^2 - 0.0468815  a^3 - 0.0131333 a^4\right. -
    0.00323908 a^5+\\&&\left. 0.00117542 a^6 \right)^{1/2},
\end{eqnarray*}
where
$a=a_s=%-1.2
-1.2455+g^q_\sigma\sigma/\sqrt{\kappa}$.

The mass $M^*_B$ of the baryon $B$ is
\begin{eqnarray*}&&M^*_N=M^*_P=3m_u,~~M_\Lambda^*=M_\Sigma^*=2m_u+m_s,~~M_\Xi^*=m_u+2m_s.
\end{eqnarray*}
According to Guichon's QMC model \cite{guichon,guichon2}, the energy density of hadronic matter is given by
\begin{eqnarray}&&
{\cal E}
={\gamma\over(2\pi)^3}\sum_B\int\d^3k\left(\sqrt{k^2+{M^*_B}^2}
+3g^q_\omega\omega+g^q_\rho\eta_Bb_3
\right)\nonumber
+{1\over2}m_\sigma^2\sigma^2-{1\over2}m_\omega^2\omega^2-{1\over2}m_\rho^2b_3^2
\\
\label{E}
\end{eqnarray}
where $\gamma=2$ is the spin multiplicity and
$g^q_\omega,~g^q_\rho $ are quark-meson coupling constants.
The baryon density, which is the source of the field $\omega$,
and the isospin density, the source of the field $b_3$, are given, respectively, by
\begin{equation}
\rho_B= {\gamma\over 2\pi^2}\sum_B \int^{k_{F_B}} k^2\d k,\quad
\rho_3={\gamma\over 2\pi^2}\sum_B
\eta_B\int^{k_{F_B}} k^2\d k,
\label{rho}
\end{equation}
with  $\eta_P=1,~\eta_N=-1,~\eta_\Lambda=\eta_{\Sigma_0}=0,~\eta_{\Sigma_+}=2,~\eta_{\Sigma_-}=-2,~\eta_{\Xi_0}=1,~\eta_{\Xi_-}=-1.$
The relation between the fields $\omega,~b_3$ and the respective sources
is obtained minimizing $\cal E$ in (\ref{E}) with respect to
$\omega,~b_3$. We find
\begin{equation}\omega={3g^q_\omega\rho_B\over m^2_\omega},\quad b_3={g^q_\rho\rho_3\over m^2_\rho}.
\label{omega}
\end{equation}
In order to describe beta decay equilibrium, the presence of electrons and muons must also be considered,
so the energy density becomes
\begin{eqnarray}&&\nonumber
{\cal
E}={\gamma\over(2\pi)^3}\left(\sum_B\int\d^3k\sqrt{k^2+{M^*_B}^2}+\sum_l\int\d^3k\sqrt{k^2+{M_l}^2} \right)\\
&&\label{CalE} +{1\over2}m_\sigma^2\sigma^2 +{1\over2}m_\omega^2\omega^2
+{1\over2}m_\rho^2b_3^2,
\end{eqnarray}
where $k_{F_l}$ and $M_l$ denote, respectively, the lepton Fermi momentum and mass.

The energy density $\cal E$ in (\ref{CalE}) should be minimized with
respect to the baryonic and the lepton Fermi momenta,  respectively, $k_{F_B}$ and $k_{F_l}$,
under constraints for the prescribed baryon number $\rho_B$,
and the charge neutrality condition,
$${\gamma\over 2\pi^2}\left(\sum_B q_B\int^{k_{F_B}} k^2\d k-
\int^{k_{F_l}}k^2\d k\right)=0,$$
with
$q_B=1$ for positively charged baryons,
$q_B=0$ for neutral baryons and
$q_B=-1$ for negatively charged baryons.
The Lagrange multiplier controlling the baryon number
is the baryon chemical potential $\mu$ and the Lagrange multiplier controlling the charge,
is denoted by $\lambda$. The Lagrange function is readily obtained.
It is the thermodynamical potential and is given by
\begin{eqnarray}
&&\Phi={\gamma\over 2\pi^2}\left(\sum_B\int^{k_{F_B}}k^2\d k\left(\sqrt{k^2+{M^*_B}^2}
-(\mu-q_B\lambda)\right)+\int^{k_{F_l}}k^2\d k\left(\sqrt{k^2+{M_l}^2}-\lambda\right)\right)\nonumber\\&&
+{1\over2}m_\sigma^2\sigma^2+{1\over2}m_\omega^2\omega^2+{1\over2}m_\rho^2b_3^2\label{Phi-e},
\end{eqnarray}
where the fields $\omega,~b_3$ are given in (\ref{omega}). Minimization of $\Phi$ with respect to $k_{F_B}$ leads to
\begin{equation}\label{Fenergy}
\sqrt{k_{F_B}^2+M^{*2}_B}
+3g_\omega^q\omega+g^q_\rho b_3\eta_B
=\mu-q_B\lambda.
\end{equation}
The quantity $\mu-q_B\lambda$ is usually referred to as the chemical potential of baryon $B$.
\color{black}
Minimization of $\Phi$ with respect to $k_{F_l}$ leads to
\begin{equation}\label{eFE}
\sqrt{k_{F_l}^2+M_l^2}=\lambda,\end{equation}
so the Lagrange multiplier $\lambda$ is usually called the lepton Fermi energy.

To summarize, in order to describe neutral matter in $\beta$ equilibrium,
we have to minimize the energy density (\ref{CalE}) with respect to the Fermi momenta $k_{F_B},k_{F_l}$ 
and $\sigma,$ for fixed baryon density and vanishing charge density.
Equivalently, the Fermi momenta $k_{F_B},~k_{F_l}$
are obtained by solving the set of simultaneous equations (\ref{Fenergy}),
(\ref{eFE}), followed by minimization with respect to $\sigma$.
In the end, it must be ensured that $\lambda$ is such that the charge density vanishes.

\section{Result and Discussions}
We start by fixing the free parameter  $\kappa$ for the Bogoliubov model. This is obtained by fitting the nucleon mass $M=939$ MeV. The desired values of nuclear matter binding energy $E_B = \varepsilon/\rho_B-ˆ' M_N =-
15.7$ MeV at saturation density, $\rho_B = 0.15$ fm$^{-ˆ'3}$ are obtained by setting $g_s^q = 3.982$
and $3g_\omega^q = g_{\omega N }$= 9.3001. The coupling $g_\rho^q = g_{\rho N }= 8.601$ is fixed so that
$E_{sym} = 32.5$ MeV at saturation density. In the Bogoliubov-QMC, the couplings of the hyperons to
the sigma meson do not need to be fixed because the effective masses of the baryons
are determined through the three quark bound. Only $x_{\omega B}$ and $x_{\rho B }$ have to be fixed. We obtain
$x_{\omega B}$ from the hyperon potentials in nuclear matter, $U_B = -(M_B^* -M_B ) + x_{\omega B} g_{\omega N} \omega_0$ for $B = \Lambda, \Sigma,$ and $\Xi$ž to be --28, 30, and --18 MeV, respectively. 
We find that $x_{\omega\Lambda}= 0.73$, $x_{\omega\Sigma} = 1.1$ and $x_{\omega \Xi}= 0.52$,
respectively for the coupling of the $\omega$-meson to the $\Lambda$,
the $\Sigma^{\pm,0}$ and the $\Xi^{-,0}$. It is worth mentioning here that the binding of
$\Lambda$ to symmetric nuclear matter is quite well settled
experimentally, although it can vary within $\sim -31\pm 3$ MeV \cite{Fortin17} , while
the binding values of $\Sigma^-$ and $\Xi^-$ still have large
uncertainties. For the $\Sigma$-hyperon it is supposed that the
potential is repulsive because no $\Sigma$-hypernucleus has been
measured. The value +30 MeV that has been considered is only
indicative and it should be taken with care. In fact, it gives origin
to a value of $x_{\omega\Sigma}$ just above 1 that may be considered
too large. Taking a smaller value of $U_\Sigma$ would decrease
$x_{\omega\Sigma}$ but the overall results would not change.
The presently existing experimental results for the $\Xi$-hypernuclei seem to indicate that
the hyperon potential at 2/3 to 1 $\rho_0$, where $\rho_0$ is the
saturation density, is approximately −14 MeV \cite{xi}. We have considered
$U_\Xi(\rho_0)=-18$ MeV, a value frequently taken in the
literature. For the $\rho$-meson-hyperon coupling we consider $x_{\rho
  B }= 1$, and the relative strength for each species is defined by
the isospin component, in particular it does not couple to the hyperons
$\Lambda$, $\Sigma^0$ and $\Xi^0$.

\begin{figure}[h]
\centering
\includegraphics[width=.5\textwidth, height=0.4\textwidth]
{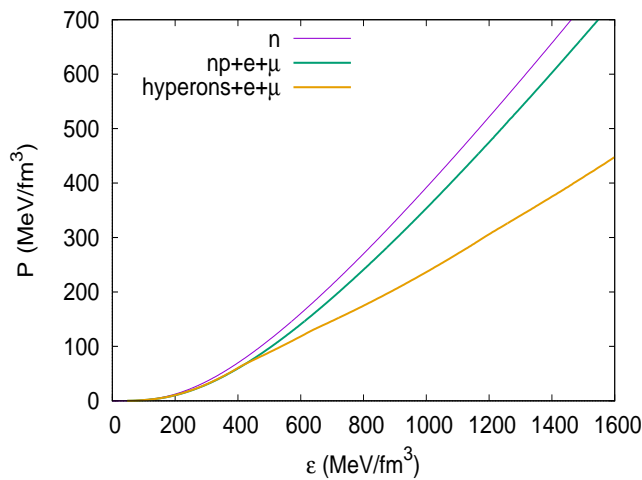} \caption{EoS for pure neutron matter and for nucleonic
  matter and hyperonic matter in $\beta$-equilibrium with  electrons and muons.      }
\label{fig07}
\end{figure}

In figure (\ref{fig07}), we show the equation of state for the pure neutron matter, neutron-proton
in $\beta$-equilibrium and the hyperon matter in $\beta$-beta
equilibrium. As expected, the neutron EoS is the hardest one and the
hyperonic EoS the softest one.

\begin{figure}[h]
\centering
\includegraphics[width=.5\textwidth, height=0.4\textwidth]
{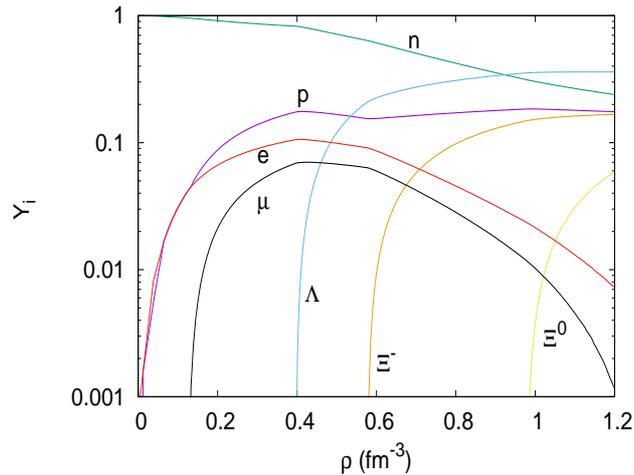} \caption{Baryonic and leptonic particle fractions
as a function of the baryonic density. At high densities, the
numbers of leptons present are small.
}
\label{fig08}
\end{figure}
In figure (\ref{fig08}) we have plotted the baryonic and the leptonic
particle  fractions.  As in other models, that take similar potentials
for the hyperons in symmetric matter, the $\Lambda$-hyperon is the
first hyperon to set in and the $\Xi^-$ the second one
\cite{cavagnoli11}. In the present model $\Xi^0$ is the third hyperon
to set in but this is not always the case as shown in
\cite{fortin16}. At  high density, the hyperonic content is influenced
if the  mesons with hidden strangeness as $\sigma^*$  and
$\phi$ are also included in the model \cite{debarati12,Fortin17}. Due
to the large uncertainties with respect to fixing their couplings, as
information on double hyperon nuclei is residual, in the present study
we do not consider them.

\begin{figure}[ht]
\centering
\includegraphics[width=.5\textwidth, height=0.4\textwidth]
{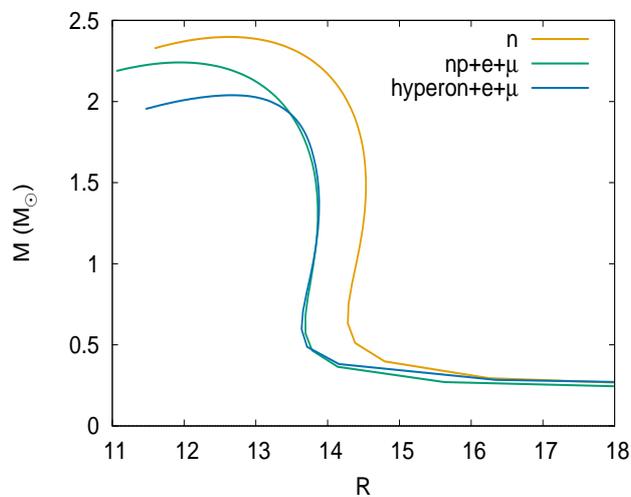} \caption{ Neutron star mass versus the radius for neutron
matter EoS and the  $\beta$-equilibrium nucleonic and hyperonic EoS,
all three EoS obtained for the  Boguliobov-QMC model.}
\label{fig01}
\end{figure}
\begin{table}[h]
\caption{Neutron star properties obtained from the integration of the
  TOV equations, maximum gravitational  and baryonic masses and
  respective radius and central energy density, radius of a $1.4M_\odot$ star and }
\begin{tabular}{lcccccc}
     &       &       &           &          \\
\hline
&~~ $M(M_\odot)$~~& ~~$M_b(M_\odot)$~~& ~~R (km)~~ &
~~$\varepsilon_c$(fm$^4$)~~& ~~$R_{1.4}$ (km)  \\
\hline
     &       &       &           &       &   & \\
n matter      & 2.39     & 2.83 & 12.63   & 5.08 &14.54   \\
np+e+$\mu$ matter     & 2.24    & 2.63 & 11.95  & 5.70  &13.86    \\
hyperon+e+$\mu$ matter   & 2.03    & 2.35 & 12.67 & 4.81 &13.86 \\
\hline
\end{tabular}
\label{star:properties}
\end{table}
We next proceed to calculate the properties of neutron  stars using the Boboliubov-QMC
model. The equation of state enters as input to the TOV equation which generates
the macroscopic stellar quantities, the mass and  the radius. In Table
\ref{star:properties} the properties of the maximum mass star are
given together with the radius of the canonical star with $M=1.4
M_\odot$.  All scenarios describe a $2M_\odot$ star as imposed by the pulsars PSR J0348$+$0432
and PSR J1614$-$2230 \cite{demorest,Antoniadis,j1614a}. Considering the
radius of the 1.4 $M_\odot$ stars, they lie within the observation
data compiled in \cite{haensel} which are still not too restrictive
due to large uncertainties. Predictions for the $R_{1.4}$ were also
obtained from the recently detection the gravitational waves GW170817
from the merging of two neutron stars: $8.7\le R_{1.4}\le14.1$ km
\cite{lattimer} or  $11.82\le R_{1.4} \le 13.72$ km \cite{tuhin}. The
predictions  from our model are within the range defined in
\cite{lattimer} and not far from the upper limit obtained in \cite{tuhin}.

\subsection{Bodmer-Witten conjecture}
{ According to the ``strange matter hypothesis" of Bodmer and Witten \cite{bodmer,witten},
% E. Witten, "Cosmic Separation Of Phases" Phys. Rev. D30, 272 (1984)
% A. Bodmer "Collapsed Nuclei" Phys. Rev. D4, 1601 (1971)
when the number of quarks  is very large,
the lowest energy state is such that it has the same number  of up, down, and strange quarks.
This stability is considered to be
a consequence of the Pauli exclusion principle, because, for three types of quarks instead of two,
as in normal nuclear matter, more quarks may be placed in the lower energy levels. This hypothesis
may be extended to hyperons, rather than quarks. Indeed,
according to the  Pauli exclusion principle, it is even more advantageous to have six hyperons instead of two nucleons.}

\begin{figure}[ht]
\centering
\includegraphics[width=.5\textwidth, height=0.4\textwidth]
{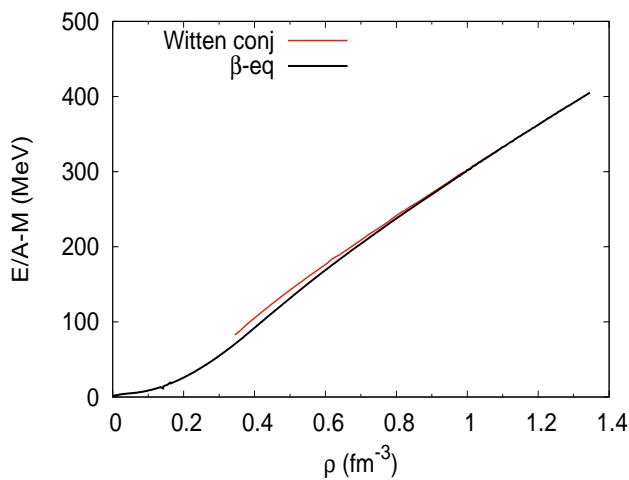} \caption{(Witten) Energy density vs. $\rho$. Approximation based on  Witten conjecture (red curve);
exact result including hyperons and beta equilibrium (black curve).}
\label{fig02}
\end{figure}

\begin{figure}[ht]
\centering
\includegraphics[width=.5\textwidth, height=0.4\textwidth]
{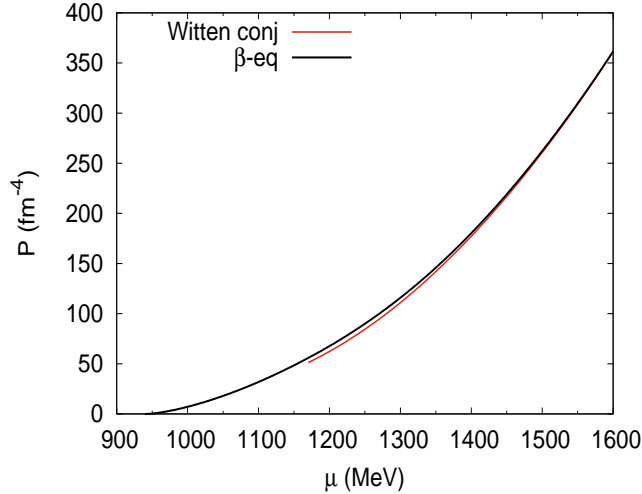} \caption{(Witten) Pressure vs. $\mu$. Approximation based on Witten conjecture (red curve);
exact result including hyperons and beta equilibrium (black curve).}
\label{fig03}
\end{figure}

In order to discuss the Bodmer-Wigner hypothesis \cite{bodmer,witten}, we replace eq. (\ref{Phi-e}) by
\begin{eqnarray*}
&&\Phi={\gamma\over(2\pi)^3}\sum_B\int^{k_{F_W}}\d^3k\left(\sqrt{k^2+{M^*_B}^2}
-\mu\right)+{1\over2}m_\sigma^2\sigma^2+{1\over2}m_\omega^2\omega^2
\label{Phi-W},
\end{eqnarray*}
where the common Fermi momentum of all baryons, $k_{F_W},$ is the solution of the following equation,
\begin{equation}\nonumber\label{wittner}\sum_B\sqrt{k_{F_W}^2+M^{*2}_B}
+3g_\omega^q\omega
=\mu.
\end{equation}
The vector field does not contribute and the source of the  $\omega$ field is given by
\begin{eqnarray*}
&&\rho= {8\gamma\over(2\pi)^3} \int^{k_{F_W}}\d^3k,
\label{rhoW}
\end{eqnarray*}
where the factor 8 accounts for the flavor degeneracy.  The Bodmer-Wigner hypothesis
is asymptotically exact in the context of the Bogoliubov independent quark model and is
a good approximation to eq. (\ref{Phi-e}) already at the center of the star densities, as Figs.
\ref{fig02} and \ref{fig03} demonstrate.

Fig. \ref{fig08} shows that for $\rho=1 fm^{-3},$ there are very few leptons.
Moreover, some hyperons are more abundant than others, depending on
their masses and charges.
What is surprising is that,
as Figs. (\ref{fig02}) and (\ref{fig03}) show, the performance of approximation based on the Bodmer-Witten
hypothesis, which ignores the difference in mass of the several hyperons,
is so good, much better than it might be expected,
at first sight. At low density, the flavor $su(2)$
symmetry prevails, but  at high densities the flavor $su(3)$ symmetry is
rather well restored.
It may also be noticed that the Bogoliubov bag is
spherical and the quarks sit at the surface. As Eq.(3) of \cite{bohr} shows,
the radius of the bag is $g^q_\sigma\sigma/\kappa$.
 It may be seen that at the neutron star  { center}, a sphere with the radius of the bag
contains about 8 nucleons. In this sense, it may be said that deconfinement
is taking place.

\section{Conclusions}
{
We have investigated a relativistic model, the Bogoliubov-QMC model, of neutral hyperonic matter in beta equilibrium in which
the quarks, up, down and strange, are considered fundamental constituents,
and hyperons are described as composite particles,  in the
framework of Bogoliubov's independent quark model. The quarks interact in the vacuum through a
linear interaction, and medium effects are taken into account
through the coupling of the quarks to mesons fields. The mesonic
fields are obtained through a minimization of the thermodynamical
potential. The parameters of the model are chosen so that saturation
nuclear matter properties are described. The size of the baryonic bags increases with density,
and for a density of about $0.8$ fm$^{-3}$ they strongly overlap, suggesting a phase
transition to quark matter.
Thus, an interesting EoS embodying the hadron-quark phase
transition may be regarded to have  been obtained.
It is found that strangeness softens the EoS and leads to a
convenient reduction of the neutron star radius. The structure of
neutron stars described within the present framework have been
calculated and it was shown that the model predicts masses above
2$M_\odot$ even if hyperons are taken into account. Also the radius of
the canonical neutron star mass with a mass equal to 1.4 $M_\odot$
comes within expected values.}

\section*{ACKNOWLEDGMENTS}

This work was supported by Funda\c c\~ao para a Ci\^encia e Tecnologia, Portugal,
under the projects UID/FIS/04564/2016 and POCI-01-0145-FEDER-029912 with  financial support
from  POCI,  in  its FEDER  component,  and  by  the FCT/MCTES  budget through  national  funds  (OE).

\section{Appendix}
The trial wave function $\Psi_{\betab,\lambda}$ of  $h^2_{D0}$
 considered in \cite{bohr} contains 2 variational parameters, $\lambda$ and $b$.
 Minimizing the expectation value of $h_{D0}^2$ with respect to $\lambda$ and $b,$
the following expression for the quark mass is found,
\begin{equation}{m^2(\kappa,\aa)\over\kappa}=\min_{\lambda,\bb}{\langle \psi_{\betab,\lambda}|h_{D0}^2|\psi_{\betab,\lambda}\rangle
\over\kappa\langle
\psi_{\betab,\lambda}|\psi_{\betab,\lambda}\rangle}=
\min_{\lambda,\bb}{{\cal K}_0+{\cal V}_0+{\cal V}_{01}\la+({\cal
K}_1+{\cal V}_1)\la^2\over {\cal N}_0+{\cal
N}_1\la^2}.\label{m^2/K}\end{equation}
Minimization of eq. (\ref{m^2/K}) w.r.t. $\lambda$ is readily
performed, so that
\begin{equation}{m^2(\kappa,\aa)\over\kappa}={1\over2}\min_\betab\left({{\cal K}_0+{\cal V}_0\over{\cal N}_0}+{{\cal K}_1+{\cal V}_1\over{\cal N}_1}
-\sqrt{\left({{\cal K}_0+{\cal V}_0\over{\cal N}_0}-{{\cal
K}_1+{\cal V}_1\over{\cal N}_1}\right)^2+\left({{\cal
V}_{01}\over\sqrt{{\cal N}_0{\cal
N}_1}}\right)^2}~\right)\label{m^2-alpha}.\end{equation}
The quantities ${\cal K}_0,~{\cal V}_0,~{\cal N}_0,~{\cal K}_1,~{\cal V}_1,~{\cal N}_1,~{\cal V}_{01}$
depend on $b$ and are given in \cite{bohr}.
Minimization of the r.h.s. of eq. (\ref{m^2-alpha}) with respect to
$\bb$ may be easily implemented. The result of this minimization is
well reproduced by (\ref{quarkmass}) in the considered interval.

In order to obtain the EoS, we need
the quark mass under the effect of an external scalar field,
$m(\kappa,g_\sigma\sigma/\sqrt\kappa)$.
%%%%%%%%%%%%%%%%%%%%%%%%%%%%%%%%%%%%%%

\end{document}